\documentclass[letter,apj]{emulateapj-rtx4}
\pdfoutput=1
\usepackage{graphicx}
\usepackage{enumerate}
\usepackage{amssymb, amsmath}
\usepackage{natbib}
\usepackage{color}
\usepackage{ulem}

\newcommand{\ias}{1}
\newcommand{\princeton}{2}
\newcommand{\sagan}{3}

\begin{document}

\title{Bayesian Ages for Early-Type Stars from Isochrones Including Rotation, and a Possible Old Age for the Hyades}
\author{Timothy D.~Brandt\altaffilmark{\ias, \sagan} \&
Chelsea X.~Huang\altaffilmark{\princeton}
}

\altaffiltext{\ias}{School of Natural Sciences, Institute for Advanced Study, Princeton, NJ, USA.}
\altaffiltext{\princeton}{Department of Astrophysical Sciences, Princeton University, Princeton, NJ, USA.}
\altaffiltext{\sagan}{NASA Sagan Fellow}

\begin{abstract}
We combine recently computed models of stellar evolution using a new treatment of rotation with a Bayesian statistical framework to constrain the ages and other properties of early-type stars.  We find good agreement for early-type stars and clusters with known young ages, including $\beta$ Pictoris, the Pleiades, and the Ursa Majoris Moving Group.  However, we derive a substantially older age for the Hyades open cluster ($750 \pm 100$ Myr compared to $625 \pm 50$ Myr).  This older age results from both the increase in main-sequence lifetime with stellar rotation and from the fact that rotating models near the main-sequence turnoff are more luminous, overlapping with slightly more massive (and shorter-lived) nonrotating ones.  
Our method uses a large grid of nonrotating models to interpolate between a much sparser rotating grid, and also includes a detailed calculation of synthetic magnitudes as a function of orientation.  We provide a web interface at \verb|www.bayesianstellarparameters.info|, where the results of our analysis may be downloaded for individual early-type ($B-V \lesssim 0.25$) {\it Hipparcos} stars.  The web interface accepts user-supplied parameters for a Gaussian metallicity prior and returns posterior probability distributions on mass, age, and orientation.
\end{abstract}

\section{Introduction}

Stellar ages are important in many areas of astrophysics, from understanding the evolution of galaxies, to chemical enrichment, to the properties of planetary systems.  They are also usually inaccessible by direct measurement and must be obtained by fitting to models \citep{Soderblom_2010}.  The one exception is the Sun, radiometrically dated from inclusions in meteorites \citep{Connelly+Bizzarro+Krot+etal_2012}.  Stars older than $\sim$100 Myr may be dated by fitting models of stellar evolution \citep{Perryman+Brown+Lebreton+etal_1998, Takeda+Ford+Sills+etal_2007} or by using calibrated empirical methods like lithium depletion \citep{Sestito+Randich_2005} or the decline of activity and rotation \citep{Noyes+Hartmann+Baliunas+etal_1984, Barnes_2007, Mamajek+Hillenbrand_2008}.  These empirical methods are calibrated at ages of several hundred Myr using clusters dated by fitting model isochrones.  

Low-mass field stars may be reliably dated by the decline of their rotation and activity.  FGK stars develop an outer convective zone, generating a magnetic field and driving a wind that carries away angular momentum \citep{Parker_1955, Glatzmaier_1985}.  They spin down and become less chromospherically and coronally active with age \citep{Wilson_1963, Hempelmann+Schmitt+Schultz+etal_1995}, providing a clock that can give ages to $\sim$20\% \citep{Mamajek+Hillenbrand_2008}.  Isochrone fits to coeval clusters are used to calibrate the empirical rotation and activity relations at ages of several hundred Myr.  The accuracy and precision of gyrochronology ages thus rely on the those of isochrone fits.  

Stars more massive than $\sim$1.3 $M_\odot$ have radiative envelopes and convective cores, and retain their angular momentum throughout the main sequence \citep{Barnes_2003}.  Even individual early-type stars are typically dated by fitting model isochrones \citep[e.g.][]{Zorec+Royer_2012, Nielsen+Liu+Wahhaj+etal_2013}.  Given an initial mass and metallicity, stellar evolution models predict the stellar radius and luminosity as a function of time.  Together with a model atmosphere, theoretical isochrones can be used to generate a grid of luminosities in each band as a function of initial mass, metallicity, and age.  A star's distance and fluxes can then be inverted to give a mass, metallicity, and age.  Unfortunately, the inversion is often not unique, with multiple paths in the color-magnitude diagram crossing one another \citep{Soderblom_2010}.  

Models of stellar evolution are subject to a very wide range of uncertainties, from the treatment of convection and overshooting, to the composition and opacities, to the treatment of rotation.  Rotation, in particular, can introduce moderate changes in luminosity, color and brightness variations with viewing angle, and large, $\sim$25\% increases in the main-sequence lifetime as rotationally-induced mixing supplies the core with fresh hydrogen \citep{Meynet+Maeder_2000, Ekstrom+Georgy+Eggenberger+etal_2012}.  The dependence of main-sequence lifetime on rotation can broaden the main-sequence and especially the main-sequence turnoff \citep{Georgy+Granda+Ekstrom+etal_2014, Li+deGrijs+Deng_2014}, mimicking the effect of a spread of ages.  The effects of rotation are most important for stars that do not brake magnetically, i.e., stars $\gtrsim$1.3 $M_\odot$.  These are precisely the stars for which isochrone dating is the most important, and often the only, way of estimating their age.  

In this paper, we perform isochrone dating of early-type stars using a recent set of stellar models with a shellular treatment of rotation \citep{Georgy+Ekstrom+Granada+etal_2013}.  These models cover a range of metallicities from $Z_\odot = 0.002$ to $Z = 0.014$, initial rotation rates from 0 to 95\% of breakup, and masses from 1.7 $M_\odot$ to 15 $M_\odot$.  We use a much finer grid of nonrotating stellar models \citep{Girardi+Bertelli+Bressan+etal_2002} to interpolate between the rotating models, and to extrapolate to $1.45 M_\odot$ and to $Z = 0.04$.  The extrapolations rely on the fact that the additional effects of rotation in these models depend only weakly on mass and metallicity.  We use a Roche model of the star \citep{Lara+Rieutord_2011} together with the ATLAS9 model atmospheres \citep{Castelli+Kurucz_2004} to compute synthetic photometry for each stellar model as a function of orientation.  We then adopt a Bayesian framework to constrain stellar parameters.

We organize our paper as follows.  In Section \ref{sec:bayesian} we describe our Bayesian statistical formalism, while in Section \ref{sec:models} we describe the stellar models that we use and the effects of adding rotation.  In Section \ref{sec:fluxcompute} we describe our use of the Roche model of a star together with model atmospheres to compute synthetic photometry.  In Section \ref{sec:results} we apply our method to four nearby early-type stars and to three coeval associations, the Pleiades, Ursa Majoris, and the Hyades.  We conclude with Section \ref{sec:conclusions}.

\section{Statistical Framework} \label{sec:bayesian}

The likelihood of a stellar evolution model is the probability of measuring the observed fluxes and distance given the predicted luminosities in each band:
\begin{equation}
{\cal L}(M, Z, \tau, \varpi) = p \big(\{m_j, \varpi_{\rm obs} \} | \{ M_j (M, Z, \tau), \varpi \} \big)~,
\label{eq:basiclike}
\end{equation}
where $M$ is the model's mass, $Z$ its metallicity and $\tau$ its age, $\varpi$ is the parallax, $m_j$ are the apparent magnitudes in bands $j$, and $M_j$ are the (absolute) model magnitudes.  The addition of rotation in stellar models enables us to include the measured projected rotational velocity, $v \sin i$, in the likelihood function.  Stellar rotation also modifies the main sequence lifetime, apparent luminosity, and colors, which all become functions of the angular velocity $\Omega$ and inclination $i$ ($i = 0$ denoting pole-on).  These become additional parameters in Equation \eqref{eq:basiclike} upon which the $M_j$ depend.  

For convenience, and for lack of an obvious alternative, we assume that the errors in observed magnitudes are Gaussian.  We neglect errors in the synthetic photometry apart from a possible systematic error that we add in quadrature to the photometric errors.  Such a systematic error is best interpreted 
as a color uncertainty.  The error in the {\it Hipparcos} parallax (which we also assume to be Gaussian) is typically a few percent for nearby stars \citep{vanLeeuwen_2007}, which is equivalent to much more than a 1\% uncertainty in flux.  With these assumptions and simplifications, the likelihood function becomes 
\begin{align}
-2 \ln {\cal L} = &\sum_{{\rm bands}\,j} \frac{\left( M_{{\rm mod},\,j}(\mu, \Omega) + 5 \log_{10} 100/\varpi - m_{{\rm obs},\,j} \right)^2}{\sigma^2_i}
\nonumber \\
&+ \frac{\left( \varpi - \varpi_{\rm obs} \right)^2 }{\sigma^2_\varpi} + \frac{(R_{\rm eq}\Omega\sin i - v_{\rm obs})^2}{\sigma^2_v} ~,
\label{eq:likefunc2}
\end{align}
where $M_{{\rm mod},\,j}$ is the model absolute magnitude in band $j$, $\varpi$ is the parallax in milliarcseconds, $R_{\rm eq} \Omega$ is the model equatorial velocity, and $v_{\rm obs}$ is the observed projected rotational velocity.

In a Bayesian framework, the posterior probability of the model is the product of the likelihood and the prior probability of the model parameters: $Z$, $M_0$, $\tau$, $\varpi$, $i$ and $\Omega$:
\begin{equation}
p(M, Z, \tau, \Omega, i) \propto p(M) p(Z) p(\varpi) p(i) p(\Omega) {\cal L}(M, Z, \tau, \Omega, i) ~.
\label{eq:postprob}
\end{equation}
The appropriate priors on $M_0$, $\tau$, $\varpi$, and $i$ are all well-established: uniform in volume and time, and proportional to the initial mass function (IMF).  The prior on parallax is thus $dp/d\varpi \propto \varpi^{-4}$, while $dp/di = \sin i$, where $i = 0$ denotes a polar viewing angle.  We adopt a Salpeter IMF, $dp/dM \propto M^{-2.35}$, appropriate for stars $\gtrsim$1 $M_\odot$ \citep{Salpeter_1955, Kroupa_2001, Chabrier_2003}.  

An appropriate prior on $Z$ in the absence of other information about the star may be taken from the metallicities of nearby young stars.  There is some disagreement on the chemical composition of the young Solar neighborhood.  The large spectroscopic sample of \cite{Casagrande+Schonrich+Asplund+etal_2011} implies a slightly sub-Solar mean metallicity for young stars, and a dispersion $\sim$0.1 dex.  This may disagree with \cite{Przybilla+Nieva+Butler_2008} and \cite{Nieva+Przybilla_2012}, who compared high resolution, high signal-to-noise ratio spectra of twenty nearby early B stars to detailed atmospheric models.  \cite{Nieva+Przybilla_2012} found abundances almost identical to those of the Solar photosphere \citep{Asplund+Grevesse+Sauval+etal_2009}, with a scatter of just $\sim$10\%, or 0.03 dex, even when including measurement uncertainties.  These results are in much better agreement with models of Galactic chemical enrichment and mixing in the interstellar medium \citep{Roy+Kunth_1995, Chiappini+Romano+Matteucci_2003}.  The data in \cite{Casagrande+Schonrich+Asplund+etal_2011} also show some trends that are unlikely to be real, including a correlation of metallicity with stellar mass at fixed age (their Figure 16).  We suggest a Gaussian prior centered on $[{\rm Fe/H}] = 0$ with a dispersion of 0.1 dex as a very conservative (perhaps overly conservative) choice.  

Our suggested metallicity prior is problematic for chemically peculiar early-type stars.  These stars are expected to have unremarkable bulk compositions, but atmospheres heavily influenced by diffusion and/or magnetic fields \citep{Preston_1974, Smith_1996}.  Individual elements and groups of elements can be over- or under-abundant by factors of tens or hundreds.  Chemically peculiar stars are relatively common on the upper main sequence, and our results for them must be interpreted with caution.  Such stars should be perhaps be fitted by a stellar model with different bulk and atmospheric compositions.

The prior on stellar rotation is more difficult to determine.  Stellar evolution codes calculate a star's structure throughout its main-sequence life, which can entail a significant redistribution of its angular momentum.  The only free parameter in this case is a star's {\it initial} supply of angular momentum, which need not be represented by its observed equatorial velocity (and which is shed in a magnetized wind by stars $\lesssim$1.3 $M_\odot$).  An appropriate prior probability on rotation would be the distribution of angular momentum (or equatorial velocity, often the only observable) for young early-type stars, which do not shed angular momentum.  

\cite{Zorec+Royer_2012} have estimated the evolution of equatorial velocity for stars in various mass bins at different stages of evolution.  For $\sim$1.5--3 $M_\odot$ stars early in their main-sequence lifetimes, \citeauthor{Zorec+Royer_2012} find peaks in the rotational velocities consistent with a bimodal distribution.  The first peak is at low velocities while the second, which accounts for most of the stars, may be approximated by a lagged Maxwellian with a mode of $\Omega/\Omega_{\rm crit} \sim 0.5$.  Their analysis also hints that $\Omega/\Omega_{\rm crit}$ may increase somewhat with initial stellar mass.

We use the models of \cite{Georgy+Ekstrom+Granada+etal_2013} in our analysis, which have equatorial $\Omega/\Omega_{\rm crit}$ somewhat lower than the initial values at which the models are tabulated.  We therefore use distributions in $\Omega/\Omega_{\rm crit}$ peaking at somewhat larger values than those derived by \cite{Zorec+Royer_2012}.  For simplicity, we use a simple Maxwellian distribution with a mode of 0.5 truncated at 0.95.  This captures the essential elements of the data \citeauthor{Zorec+Royer_2012} analyze,  apart from a few ($\sim$0--20\%, depending on mass) of stars that appear to rotate slowly, and reproduces the approximate upper limits to observed values of $v \sin i$.

\section{Stellar Evolution Models With Rotation} \label{sec:models}

A wide range of stellar evolution models is now available, covering fine grids of mass, metallicity, and time \citep{Yi+Demarque+Kim+etal_2001, Girardi+Bertelli+Bressan+etal_2002, Pietrinferni+Cassisi+Salaris+etal_2004, Dotter+Chaboyer+Jevremovic+etal_2008}.  Some models, going back more than a decade, have included rotation \citep{Meynet+Maeder_2000}.  The addition of rotation turns a 1D stellar structure into at least a 2D structure.  This is simplified in practice with assumptions like that of ``shellular'' rotation \citep{Zahn_1992}, in which the angular velocity is constant along isobars.  Rotation is relatively unimportant for stars below $\sim$1.3 $M_\odot$, which spin down rapidly.  However, it can introduce large effects for stars with radiative envelopes \citep{Meynet+Maeder_2000}, perhaps most notably an increase in time spent on the main sequence.

Stellar modeling includes a very wide range of tunable physical parameters governing everything from the chemical abundances and their opacities, to the treatment of convection, to the treatment (or lack) of rotation, to the model atmospheres.  These parameters and treatments vary from one set of stellar models to another and can lead to significant variations in colors, luminosities, and lifetimes \citep{Lebreton+Goupil+Montalban_2014}.  For the rest of this analysis, we adopt the models of \cite{Georgy+Ekstrom+Granada+etal_2013} as our primary source, compute synthetic photometry ourselves, and neglect uncertainties in parameters like the helium fraction or convective overshooting.

The recent stellar evolution models of \cite{Ekstrom+Georgy+Eggenberger+etal_2012} and \cite{Georgy+Ekstrom+Granada+etal_2013} include both rotating and non-rotating stars, in the former case covering a range of initial rotation rates and metallicities.  These grids are, however, very coarse, and only extend to $Z = Z_\odot$.  We therefore use much finer grids of nonrotating models to interpolate between the rotating models, and to extrapolate them to higher metallicities.  We adopt the PARSEC isochrones \citep{Girardi+Bertelli+Bressan+etal_2002} for this interpolation.  Because we anchor them to the \cite{Georgy+Ekstrom+Granada+etal_2013} models, the details of the PARSEC calculations are unimportant here.  Any effects that are linearly dependent on mass and metallicity will disappear entirely.  

The validity of our interpolation and extrapolation relies on the fact that the modifications to stellar evolution provided by rotation depend only weakly on other stellar parameters.  In this analysis, we restrict the models to $M > 1.5 M_\odot$, for which magnetic braking is inefficient and rotation is important throughout the main sequence.  We apply our upper mass cutoff at $11 M_\odot$, which has a main sequence lifetime of just a few tens of Myr.  

The rapidly rotating stellar models have main-sequence lifetimes that depend strongly on $\Omega$, the initial angular momentum.  Interpolating with nonrotating models requires us to introduce a dimensionless time, $t/t_{\rm MS}$, the ratio of the stellar age to its main sequence lifetime.  The parameter $t_{\rm MS}$ is a function of stellar rotation:
\begin{equation}
t_{\rm MS} = t_{\rm MS,\,nr} \beta \left( M_0, \Omega \right)~. 
\label{eq:beta}
\end{equation}
We use the PARSEC models to interpolate $t_{\rm MS,\,nr}$ between masses in the Geneva models, and separately interpolate $\beta$, which depends very weakly on mass and metallicity.  We note that the nonrotating PARSEC isochrones do differ significantly from the nonrotating Geneva models.  For example, 1.5--5 $M_\odot$ PARSEC models spend $\sim$10\% longer on the main sequence than the corresponding Geneva models we use here.

The other effects of rotation include an increase in luminosity and a flattening of the stellar surface, which make the apparent luminosity of the star a function of viewing angle.  We describe the computation of synthetic photometry in the following section.  To interpolate between models, we separate the photometry into nonrotating magnitudes and a rotating correction as a function of $\Omega$, photometric band, and inclination.  These two components are additive in units of magnitude (multiplicative in units of flux).  We then use the PARSEC models to interpolate the nonrotating synthetic photometry between Geneva models.  We separately interpolate or extrapolate the rotation correction term, which, like $\beta$ from Equation \eqref{eq:beta}, depends only weakly on stellar mass and metallicity.  This interpolation in linear in units of magnitude.

\section{Computing the Effect of Orientation} \label{sec:fluxcompute}

Stellar oblateness introduces two effects: a viewer along the pole sees a star both larger in area and hotter.  In all bands, the star will therefore appear more luminous with decreasing inclination $i$ ($i = 0$ denoting a line-of-sight parallel to the polar axis), except when $i$ approaches $0^\circ$ and the gravity darkening effect is severe. The variation of stellar surface brightness with latitude is a consequence of the von Zeipel theorem \citep{vonZeipel_1924}, which states that the local effective temperature depends on the local effective gravity as a power-law $T_{\rm eff}\propto\,g_{\rm eff}^{\beta}$, with $\beta$ known as the gravity darkening coefficient. The original von Zeipel law states $\beta=1/4$ for stars with fully radiative envelopes. Generally, $\beta$ is smaller than $1/4$, as discovered by recent interferometric observations of rapidly rotating nearby stars \citep{Aufdenberg+Merand+Foresto+etal_2006,van_Belle+Ciardi+ten_Brummelaar+etal_2006}. 

Although comprehensive calculations of $\beta$ were carried out by \citet{Claret_1998} for stars at different evolutionary stages, we choose to adopt a two-dimensional model developed by \citet{Lara+Rieutord_2011} (hereafter LR11), in which the effective temperature profile only depends on the ratio of the equatorial velocity and the Keplerian velocity $w=v/v_{\rm crit}$.  This parameter is supplied by the stellar evolution models \citep{Ekstrom+Georgy+Eggenberger+etal_2012}.  With the total luminosity, oblateness, and radius from the stellar models, we can compute $T_{\rm eff}$ and $g_{\rm eff}$ over the entire stellar surface. Finally, we use the specific intensities from the ATLAS9 models \citep{Castelli+Kurucz_2004} to integrate the apparent flux through a series of filters as a function of viewing angle.  

\subsection{Computing Fluxes}

We follow Section 2 of LR11 to compute the effective temperature map over the stellar surface. The assumption of the model can be simply stated as the following: the mass distribution of the star may be described as a Roche model, and the energy flux is a divergence-free vector that is almost anti-parallel to the effective gravity. The effective temperature profile at a particular polar angle $\theta$ can then be expressed as 
\begin{align}
T_{\rm eff} = &\left( \frac{1}{r^4}+w^4r^2\sin^2\theta-\frac{2w^2\sin^2\theta}{r} \right)^{1/8} \nonumber \\
&\quad \times \left( \frac{L}{4\pi\sigma\,R_{\rm eq}^2} \right)^{1/4} 
F_w^{1/4},
\end{align}
in which $r$ is the radius at the stellar surface in units of the equatorial radius $R_{\rm eq}$, $w$ is the ratio between the equatorial velocity and the Keplerian velocity $v/v_{\rm crit}$, and $F_w$ is a correction factor that takes into account the difference between the vector angle of energy flux and the effective gravity. When the rotation is slow ($R_{\rm pole}/R_{\rm eq}>0.95$), $F_w=1$, and this model reduces to the von Zeipel case. 

We solve for $F_w$ numerically using a combined Newton-Raphson and bisection method following Equations (24)--(28) of LR11. The stellar surface is assumed to be an oblate ellipsoid with $a=R_{\rm eq}$ and $c=R_{\rm pol}$. The projection of the stellar surface onto the plane of the sky with an inclination $i$ is an ellipse with $a=R_{\rm eq}$ and $b=(R_{\rm eq}^2R_{\rm pol}^2)/(R_{\rm eq}^2\cos^2i+R_{\rm pol}^2\sin^2i)$.

We use the ATLAS9 stellar atmosphere models \citep{Castelli+Kurucz_2004} to compute the specific intensities, and the Tycho, 2MASS \citep{Cutri+Skrutskie+vanDyk+etal_2003}, and Johnson-Cousins filter response curves to compute photon fluxes at a distance of 10 pc.  We adopt the recently revised {\it Tycho} responses \citep{Bessell+Murphy_2012}, which differ slightly (by $\lesssim$5 mmag in $B_T-V_T$) from those originally published \citep{ESA_1997}.  The abundances in the ATLAS9 models are relative to  \cite{Anders+Grevesse_1989}, which gives a Solar metallicity $Z = 0.019$.  This is higher by 0.10--0.13 dex than the $Z_\odot$ values adopted by \cite{Girardi+Bertelli+Bressan+etal_2002} and \cite{Ekstrom+Georgy+Eggenberger+etal_2012} based on the abundances measured by \cite{Asplund+Grevesse+Sauval+etal_2009}.  We therefore use the ATLAS9 models with $[{\rm Fe/H}]=-0.1$ for the isochrone models at solar metallicity.  

Our adoption of $Z_\odot \approx 0.014$ could introduce systematic effects when using spectroscopic metallicities that assume a different $Z_\odot$.  As a result, $[{\rm Fe/H}] = 0.1$ in this work could correspond in some ways to a spectroscopic $[{\rm Fe/H}] = 0.0$.  Our scaling of the \cite{Anders+Grevesse_1989} photospheric composition does, however, mean that the composition of the stellar atmosphere roughly matches the bulk composition of the star.  

\subsection{A Parametrized Fit}

Our method allows us to compute the magnitudes in all bands as a function of orientation.  We wish to use the fine grid of nonrotating PARSEC isochrones \citep{Girardi+Bertelli+Bressan+etal_2002} to interpolate within the coarse grid of rotating Geneva models.  We therefore compute the difference in magnitude between the rapidly rotating models of \cite{Georgy+Ekstrom+Granada+etal_2013} and PARSEC models at the same mass, metallicity, and fraction of main sequence lifetime.  Some of the \cite{Georgy+Ekstrom+Granada+etal_2013} models are computed with zero initial angular momentum.  These models are not identical to the corresponding PARSEC models; for example, nonrotating PARSEC $\sim$2 $M_\odot$ stars typically spend $\sim$10\% longer on the main sequence than the corresponding nonrotating Geneva models.  Even at zero rotation, the difference between Geneva and PARSEC model magnitudes can be significant.

\begin{figure*}
\includegraphics[width=\textwidth]{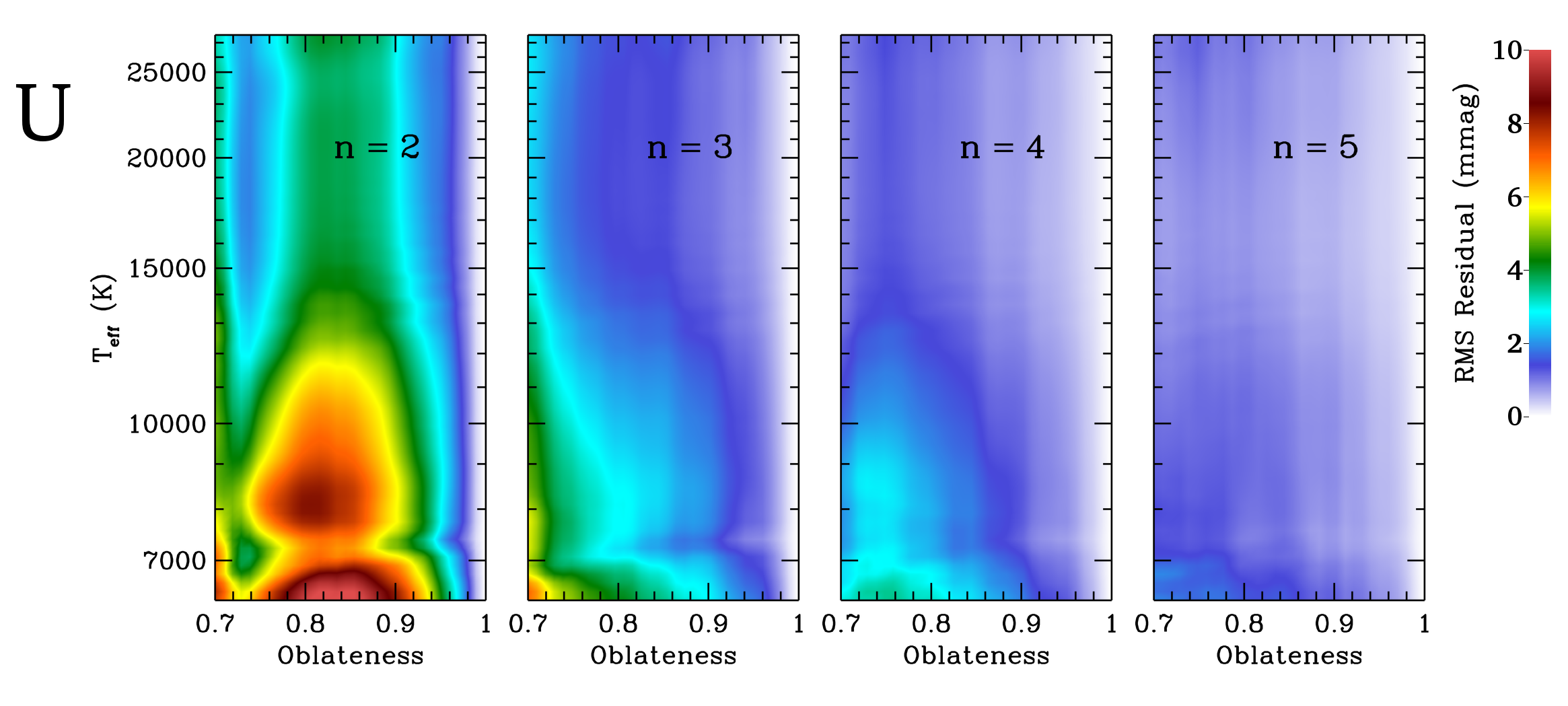}
\caption{Distribution of the residuals from Equation \eqref{eq:fittingfunc} for the Johnson-Cousins $U$-band.  With a fitting polynomial of order $n = 5$, the root-mean-square residuals are $\lesssim$1 mmag for all but the lowest temperature, most rapidly rotating stars.  This is much less than pole-equator differences reaching 0.1 mag or more, and less than errors due to interpolation and measurement errors.  The residuals are smaller still (i.e.~the fit is better) at the longer-wavelength passbands that we use.  }
\label{fig:residuals}
\end{figure*}

We compute the magnitude corrections as a function of viewing angle $\mu \equiv \cos i$ ($\mu = 1$ corresponding to pole-on).  We then fit a polynomial in $\mu$ for each band and rotating model,
\begin{equation}
M(\mu \equiv \cos i) = M_{\rm PARSEC} + \sum_{i=0}^n a_i \mu^i~. 
\label{eq:fittingfunc}
\end{equation}
Figure \ref{fig:residuals} shows the residuals from this fitting function for the Johnson-Cousins U passband as a function of $T_{\rm eff}$ and obliquity at an equatorial $\log g = 4$.  We obtain excellent agreement with $n = 5$, with residuals $\lesssim$1 mmag for all but the coolest models at the fastest rotations.  These residuals decrease towards longer-wavelength passbands.  With $n = 5$, they are typically at least an order of magnitude smaller than the photometric measurement errors, and likely smaller than errors resulting from interpolation between isochrones.  

The $a_i$ in Equation \eqref{eq:fittingfunc} are functions of the band, stellar mass, rotation rate, metallicity, and age, and are logarithmic in the stellar flux.  We linearly interpolate (and even extrapolate, in metallicity), these parameters onto the fine nonrotating PARSEC grids.  In doing so, we take advantage of the fact that while the stellar models themselves may depend strongly on physical parameters like mass and metallicity, the $a_i$ correction coefficients have a much weaker dependence.  

\begin{figure}
\includegraphics[width=0.5\textwidth]{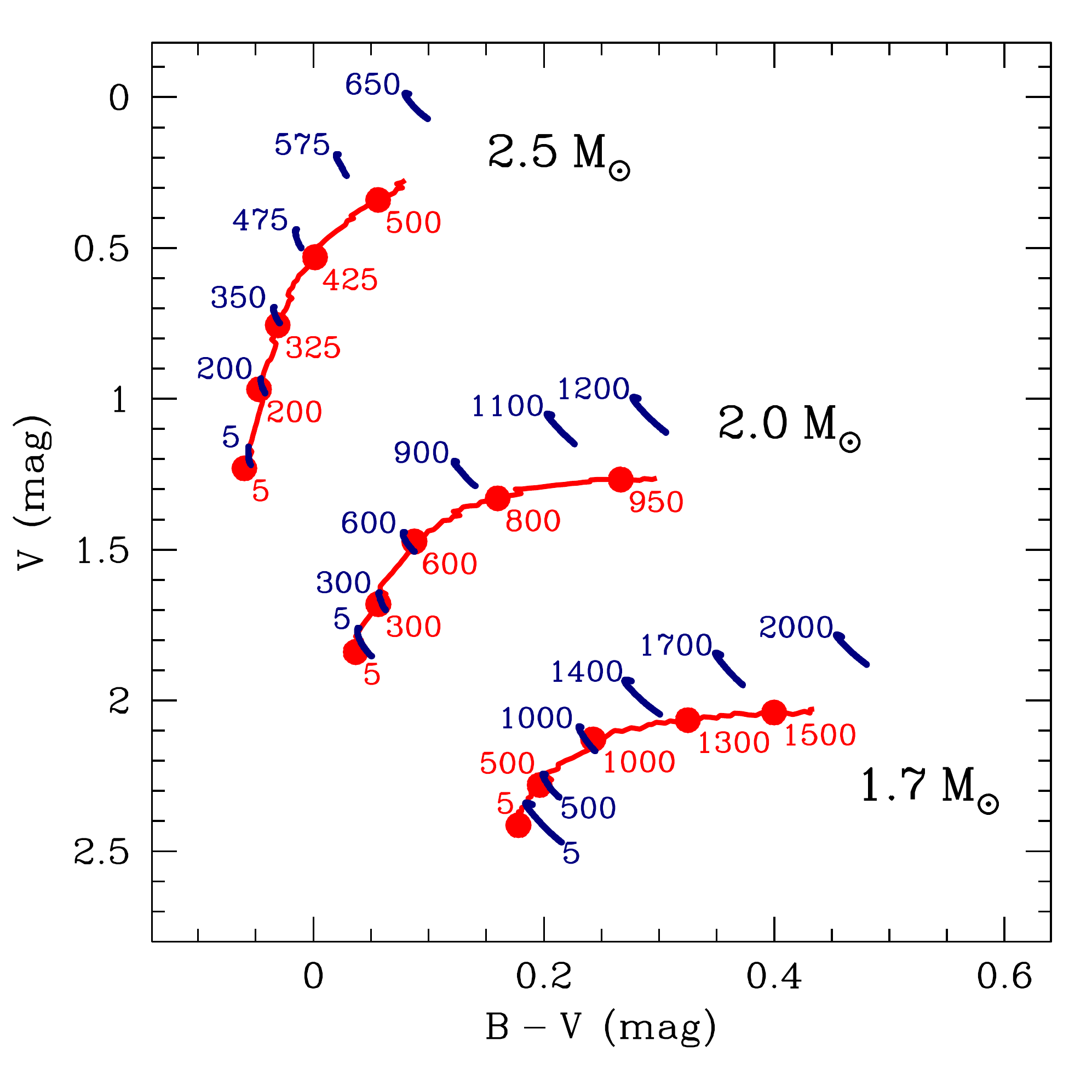}
\caption{Isochrones for three stellar masses, with (blue) and without (red) rotation at fixed $\Omega/\Omega_{\rm crit} = 0.5$; several ages are noted in Myr.  The rotating models have been plotted only at representative ages to show the effects of orientation.  The lines show the positions in color-magnitude space as a function of orientation.  At older ages, a rotating model will overlap a significantly younger, slightly more massive nonrotating star in color-magnitude space.}
\label{fig:isochrones}
\end{figure}

With the effect of rotation reduced to the parameters of Equation \eqref{eq:fittingfunc}, we can compute the effects of both rotation and orientation on observed colors and magnitudes.  Figure \ref{fig:isochrones} shows both nonrotating and rotating isochrones (with $\Omega/\Omega_{\rm crit}=0.5$) at a series of representative ages.  A rotating model at fixed age actually covers a line in the color-magnitude diagram, as much as $\sim$0.1--0.2 magntiudes in $V$ and 0.05 magnitudes in $B-V$ (the navy blue lines in the figure).  At very young ages, the effect of rotation is modest.  At older ages, however, a rotating stellar model will overlap a significantly younger and slightly more massive model in color-magnitude space.  The combination of an increased luminosity at a similar $B-V$ color with a longer time spent on the main sequence can have a very large effect on isochrone ages using main-sequence turnoff stars.  This is especially apparent for the Hyades open cluster, which we discuss at the end of the next section.

\section{Results} \label{sec:results}

In the following subsections we present the results of our isochrone analysis for a selection of A and B stars.  We apply it first to a series of three individual stars of known age and to one star, $\kappa$ And, with an imaged substellar companion and a controversial age.  We have generally excluded close visual and spectroscopic binaries but have not excluded any stars based on the performance of our fitting.  We then apply our isochrone-based analysis to three stellar clusters with previously inferred isochrone ages, the Pleiades, the Ursa Majoris moving group, and the Hyades open cluster.  

We restrict our analyses to the {\it Tycho-2} $B_T$ and $V_T$ photometry \citep{Hog+Fabricius+Makarov+etal_2000}.  These data have been well-calibrated and offer precisions of $\sim$0.01 mag down to $V_T \sim 8$ over the entire sky.  2MASS $JHK_s$ photometry could also be used, but would require a careful analysis demonstrating its calibration relative to {\it Tycho} at levels of $\lesssim$0.01 mag, and a demonstration that the model atmospheres are adequate over such a wide range of wavelengths.  We defer such an investigation to a future paper.  We use {\it Hipparcos} parallaxes throughout  \citep{vanLeeuwen_2007}, and use $v \sin i$ measurements collected and calibrated by \cite{Zorec+Royer_2012}.

In all of our fitting, we apply floors on the measurement errors of all parameters, which we add in quadrature with the actual reported uncertainties.  These error floors are 0.005 magnitudes in all bands ($\sim$0.5\% in flux), which are best thought of as color errors, and 30 km\,s$^{-1}$ in $v \sin i$.  They are designed to minimize the impact of the coarseness of our model grid and to account for some uncertainties in the synthetic photometry.  The Geneva models are only computed at nine initial angular momenta; the spacing between them corresponds to $\sim$20--50 km\,s$^{-1}$ in equatorial velocity.  The spacing between neighboring models in luminosity is typically $\sim$2\% (and typically much less in color).  The difference in synthetic colors between the original and recalibrated {\it Tycho} photometric system produces a systematic shift in $B_T-V_T$ of $\sim$0.004 mag, while errors in parallax are $\gtrsim$2\%.  

When fitting an ensemble of stars that share an age and chemical composition, we multiply the two-dimensional posterior probability distributions for each individual star.  This relies on the stellar models providing adequate fits to all stars, and on these fits being consistent.  A rigorous statistical statement of consistency depends on the actual distributions.  Roughly speaking, however, consistency requires that the $1\sigma$ contour of the multiplied posterior distribution be at a level comparable to the product of the $1\sigma$ contours in each individual distribution.  All of our cluster probability distributions presented below easily pass this test.

\subsection{Single Stars}

\begin{figure}
\centering
\includegraphics[width=\linewidth]{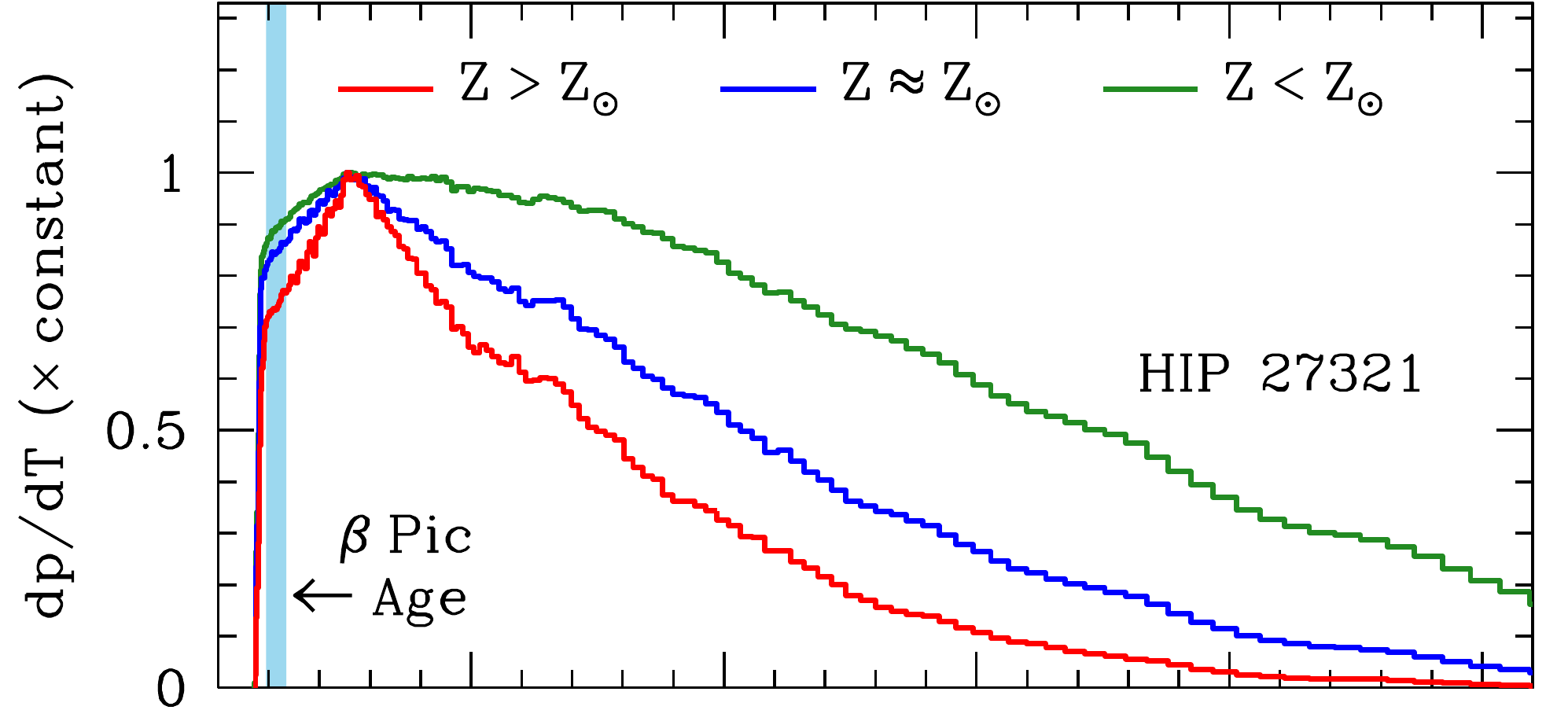}
\includegraphics[width=\linewidth]{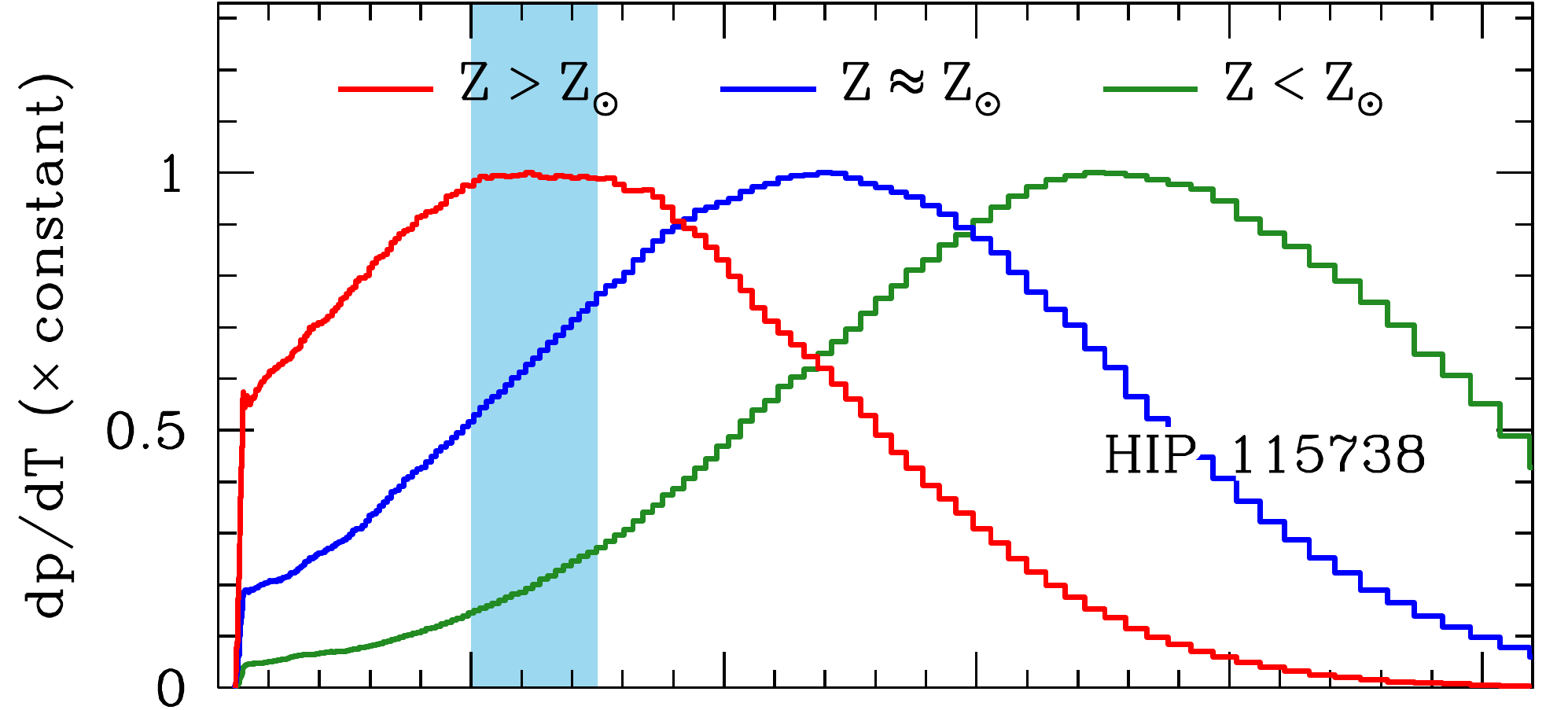}
\includegraphics[width=\linewidth]{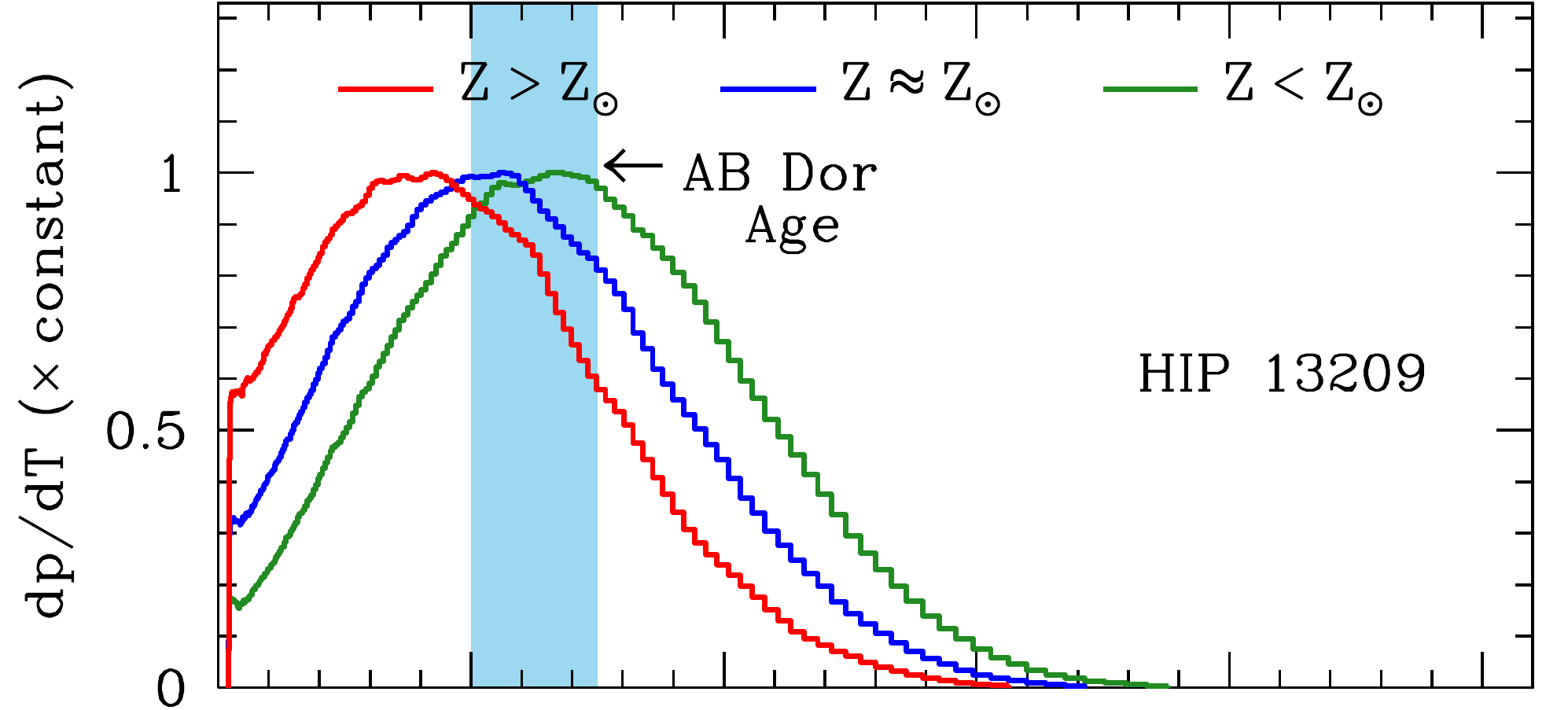}
\includegraphics[width=\linewidth]{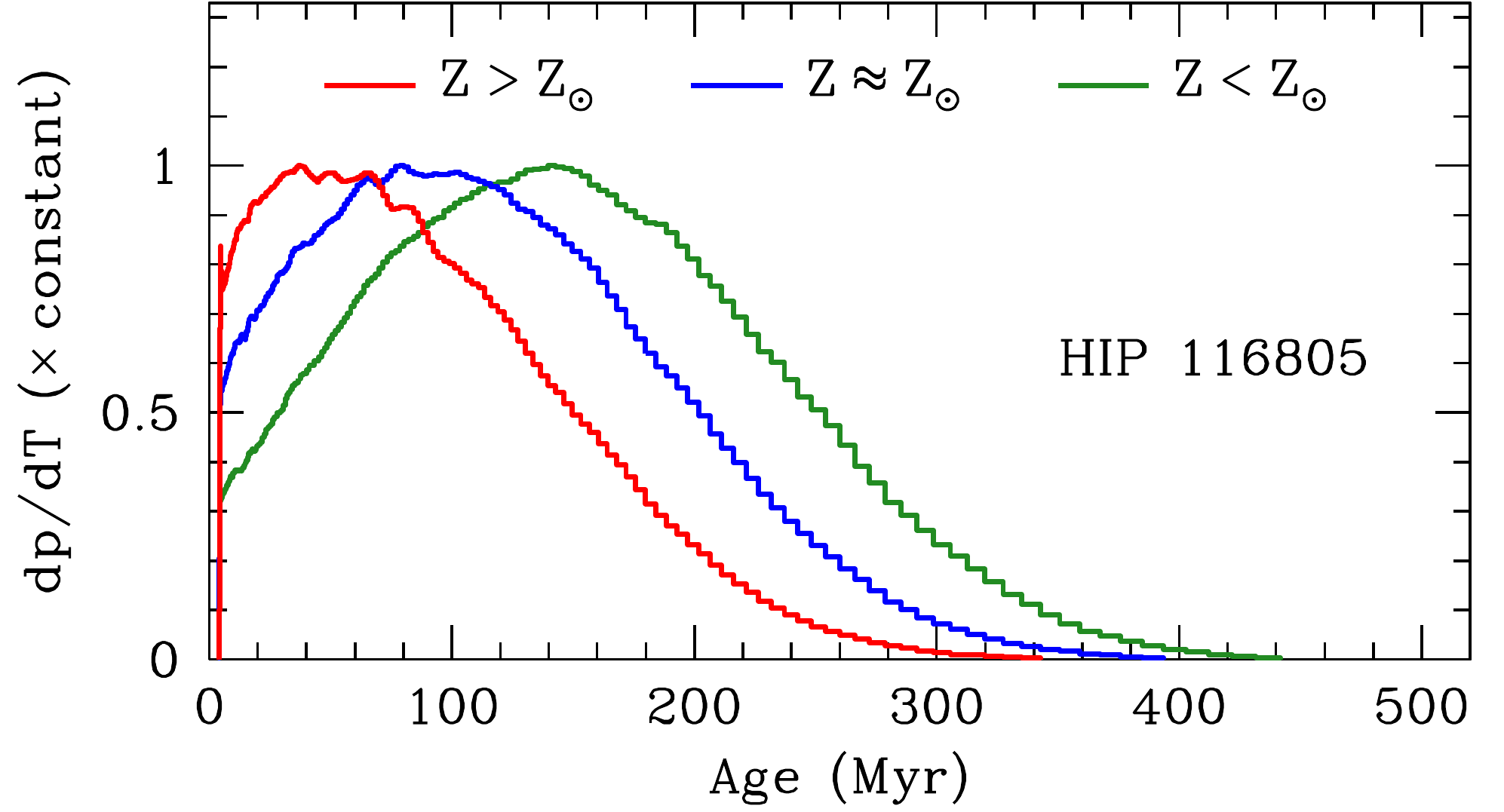}
\caption{Posterior age probability distributions for four {\it Hipparcos} stars, each with three metallicity priors: $[{\rm Fe/H}] = 0.1 \pm 0.05$ (red histograms), $[{\rm Fe/H}] = 0 \pm 0.05$ (blue histograms), and $[{\rm Fe/H}] = -0.1 \pm 0.05$ (green histograms).  HIP 13209 (third panel) has a binary companion of unknown spectral type which we neglect in the isochrone fit.  The shaded light blue regions show the moving group ages of the top three stars.  Slightly super-Solar metallicities allow an excellent match in all cases.  HIP 116805's age is controversial; our analysis suggests that youth cannot be ruled out.  }
\label{fig:singlestar_ages}
\end{figure}

We first apply our isochrone-based analysis to four individual stars: HIP 27321 ($\beta$ Pic), 13209, 115738, and 116805 ($\kappa$ And).  The first three of these stars are consensus members of the coeval moving groups $\beta$ Pictoris (HIP 27321) and AB Doradus (HIP 13209 and 115738).  HIP 116805 is a proposed member of the Columba moving group, but this identification, and the star's age, are controversial.  Figure \ref{fig:singlestar_ages} shows the age posterior probability distributions for each of these stars for three Gaussian metallicity distributions: $[{\rm Fe/H}] = 0.1 \pm 0.05$ (red histograms), $[{\rm Fe/H}] = 0 \pm 0.05$ (blue histograms), and $[{\rm Fe/H}] = -0.1 \pm 0.05$ (green histograms).  The shaded blue regions show the moving group ages.  We discuss each star in turn below.   

{\it HIP 27321}: 
The nearby A star $\beta$ Pictoris hosts a debris disk and a low-mass substellar companion \citep{Lagrange+Gratadour+Chauvin+etal_2009}.  It is also the founding member of the $\beta$ Pictoris moving group \citep{BarradoyNavascues+Stauffer+Song+etal_1999, Zuckerman+Song+Bessell+etal_2001}, whose age has recently been determined to be $\sim$20--25 Myr using the lithium depletion boundary, isochrones with magnetic fields, and kinematics \citep{Binks+Jeffries_2014, Malo+Doyon+Feiden+etal_2014, Mamajek+Bell_2014}.  

HIP 27321 has a measured rotational velocity of $\sim$120--130 km\,s$^{-1}$ \citep{Royer+Zorec+Gomez_2007, Schroder+Reiners+Schmitt_2009} and, while measurements of its metallicity are unreliable, we can adopt the chemical composition of lower-temperature members of the same moving group.  HIP 10679 has a spectroscopic $[{\rm Fe/H}] = 0.07 \pm 0.03$, HIP 10680 has $[{\rm Fe/H}] = 0.09 \pm 0.03$, and HIP 25486 has $[{\rm Fe/H}] = 0.29 \pm 0.03$ \citep{Valenti+Fischer_2005}.  As members of the same moving group, these stars should have nearly identical compositions.  The inconsistency between their spectroscopic metallicities in the same survey could indicate large systematic errors.  The measurements do, however, hint at a slightly super-solar $[{\rm Fe/H}]$ for $\beta$ Pic itself.

The top panel of Figure \ref{fig:singlestar_ages} shows the age posterior probability distribution for HIP 27321 under three metallicity priors.  The red histogram, $[{\rm Fe/H}] = 0.1 \pm 0.05$, comes closest to the metallicities determined for later-type $\beta$ Pic members, and correctly indicates a young age for HIP 27321 itself.  The lower limit on the star's age is an artifact of our masking of the pre-main sequence.  

{\it HIP 115738:} 
HIP 115738 is an $\alpha^2$ CVn variable star and a high-probability member of the AB Doradus moving group \citep{Zuckerman+Rhee+Song+etal_2011, Malo+Doyon+Lafreniere+etal_2013, Gagne+Lafreniere+Doyon+etal_2014}, with an age of $\sim$100--150 Myr \citep{Luhman+Stauffer+Mamajek_2005, Ortega+Jilinski+delaReza+etal_2007, Barenfeld+Bubar+Mamajek+etal_2013}.  The metallicity of AB Dor is somewhat uncertain, with measurements of $[{\rm Fe/H}] = 0.02 \pm 0.02$ \citep{Barenfeld+Bubar+Mamajek+etal_2013} and $[{\rm Fe/H}] = 0.10 \pm 0.03$ \citep{Biazzo+DOrazi+Desidera+etal_2012}, from spectroscopy of later-type members.  

Variable stars of $\alpha^2$ CVn type are chemically peculiar with strong metal lines.  This could imply a difference between the metallicity of the star and its atmosphere, an effect which we ignore.  While HIP 115738's variability is a generic problem for any isochrone analyses, the {\it Tycho} photometry is from a stack of 47 measurements (with a root-mean-square scatter in $V_T$ of 0.029 mag).  Stellar variability would be a much larger concern if we were to include non-simultaneous photometry at other wavelengths.  

HIP 115738's variability does, in principle, permit a measurement of the its rotational period.  Unfortunately, there seems to be little agreement on the star's photometric period among variability surveys, ranging from 2 hours \citep{Rimoldini+Dubath+Suveges+etal_2012} to 1.4 days \citep{Wraight+Fossati+Netopil_2012}.  A rotation period of 2 hours is physically impossible for an A star of $\sim$2 $R_\odot$.  An undisputed measurement of the stellar rotation period would enable us to directly constrain $\Omega$, the surface angular velocity.

The second panel of Figure \ref{fig:singlestar_ages} shows the results of our analysis for HIP 115738.  At a slightly super-Solar metallicity, our posterior probability distribution is in excellent agreement with the age of the AB Dor moving group.  HIP 115738's age distribution would also skew somewhat younger if the star is a rapid rotator seen pole-on.

{\it HIP 13209:}
The late B star HIP 13209 (41 Ari) is a consensus member of AB Dor \citep{Zuckerman+Rhee+Song+etal_2011, Malo+Doyon+Lafreniere+etal_2013, Gagne+Lafreniere+Doyon+etal_2014}.  The star is a known spectroscopic binary and has been resolved at an angular separation of $\sim$$0.\!\!''1$ \citep{McAlister+Hartkopf+Hutter+etal_1987}.  The binary companion lacks a published spectral type or contrast, and was not seen in near-infrared adaptive optics imaging \citep{Roberts+Turner+tenBrummelaar_2007, Janson+Bonavita+Klahr+etal_2011}.  HIP 13209 has a measured rotational velocity of 175 km\,s$^{-1}$ \citep{Abt+Levato+Grosso_2002}.  The metallicity of AB Dor is somewhat uncertain, with measurements of $[{\rm Fe/H}] = 0.02 \pm 0.02$ \citep{Barenfeld+Bubar+Mamajek+etal_2013} and $[{\rm Fe/H}] = 0.10 \pm 0.03$ \citep{Biazzo+DOrazi+Desidera+etal_2012}, from spectroscopy for later-type members.  

HIP 13209 is the earliest-type member of AB Dor listed in \cite{Malo+Doyon+Lafreniere+etal_2013} and we include it here in spite of its companion.  
We make no attempt to remove the companion's flux, but note that any later-type companion would tend to bias our probability distributions to slightly older ages, i.e., make the star younger than our analysis implies.  A G-type or later binary would contribute too little flux to make much difference.

The third panel of Figure \ref{fig:singlestar_ages} shows the results for our three metallicity priors.  With a Solar of slightly super-Solar metallicity, and neglecting the companion, 
our isochrone-based analysis agrees well with AB Dor's known age.

{\it HIP 116805}:
The late-B star $\kappa$ And has been observed to host a substellar companion \citep{Carson+Thalmann+Janson+etal_2013}.  The companion's mass depends strongly on the host star's age, which has been estimated from $\sim$30 to $\sim$300 Myr \citep{Carson+Thalmann+Janson+etal_2013, Hinkley+Pueyo+Faherty+etal_2013, Bonnefoy+Currie+Marleau+etal_2014}.  The older age estimates are based on fitting nonrotating isochrones to the star's position in color-luminosity space while the young age estimate relies on the star's proposed membership in the Columba moving group \citep{Malo+Doyon+Lafreniere+etal_2013}.  We revisit the stellar parameters using the formalism we develop in this paper.  

We note that $\kappa$ And does have a spectroscopic metallicity measurement giving $[{\rm Fe/H}] = -0.32 \pm 0.15$ \citep{Wu+Singh+Prugniel+etal_2011}.  However, such a young, metal-poor star would be exceptional in the Solar neighborhood.  Analyses of young clusters and moving groups, and of young stars in the local field, consistently find metallicities centered on or near the Solar value, with a dispersion $\sim$0.1 dex \citep[e.g.][]{Gaidos+Gonzalez_2002, Biazzo+Randich+Palla+etal_2011, Biazzo+DOrazi+Desidera+etal_2012}.  We therefore use the same three metallicity priors as for the other stars discussed above: $[{\rm Fe/H}] = 0.1 \pm 0.05$, $[{\rm Fe/H}] = 0.0 \pm 0.05$, and $[{\rm Fe/H}] = -0.1 \pm 0.05$.  The star has a rotational $v \sin i \approx 160$ km\,s$^{-1}$ \citep{Zorec+Royer_2012}.

The bottom panel of Figure \ref{fig:singlestar_ages} shows our results for HIP 116805.  A young age is excluded only if $\kappa$ And's metallicity is strongly sub-Solar.  If its metallicity is approximately Solar or slightly super-Solar, HIP 116805's position in color-magnitude space appears to be consistent with membership in Columba and a $\sim$30 Myr age.

\subsection{The Pleiades Open Cluster} \label{subsec:pleiades}

The Pleiades open cluster is a large and extensively studied nearby stellar association.  The Pleiades' age is $130 \pm 20$ Myr, as measured from the lithium depletion boundary \citep{BarradoyNavascues+Stauffer+Jayawardhana_2004}.  It has also been estimated at 100 Myr using stellar isochrones without rotation \citep{Meynet+Mermilliod+Maeder_1993}.  Reddening for the bulk of the cluster is $E(B-V) = 0.04$ mags, with a slight variation from star to star \citep{Vandenberg+Poll_1989, Taylor_2008}.  The distance to the Pleiades has been more controversial, but multiple lines of evidence, including several very precise parallax measurements, now point to $\sim$135 pc \citep{Pinsonneault+Stauffer+Soderblom+etal_1998, Pan+Shao+Kulkarni_2004, Soderblom+Nelan+Benedict+etal_2005, Melis+Reid+Mioduszewski+etal_2014}.  As a large, well-studied cluster, the Pleiades presents an important test of our statistical approach using the \cite{Georgy+Ekstrom+Granada+etal_2013} isochrones.  

We restrict our analysis to late B Pleiads ($\sim$2.5 $M_\odot$), HIP 17527, 17588, 17664, 17776, 17862, and 17900.  These are similar in mass to the stars that we will later use in analyses of the Hyades open cluster and the Ursa Majoris moving group.  They have evolved sufficiently in $\sim$130 Myr to make an isochrone-based analysis possible, but are relatively far from the main-sequence turnoff.  We adopt a parallax of $7.4 \pm 0.2$ mas ($135 \pm 4$ pc) and a reddening of $E(B-V) = 0.04$ mag, with $R_V = 3.1$, for all stars.  We use a metallicity of $[{\rm Fe/H}] = 0.03 \pm 0.05$, as measured spectroscopically for 20 roughly Solar-type members \citep{Soderblom+Laskar+Valenti+etal_2009}.  

\begin{figure}
\centering
\includegraphics[width=\linewidth]{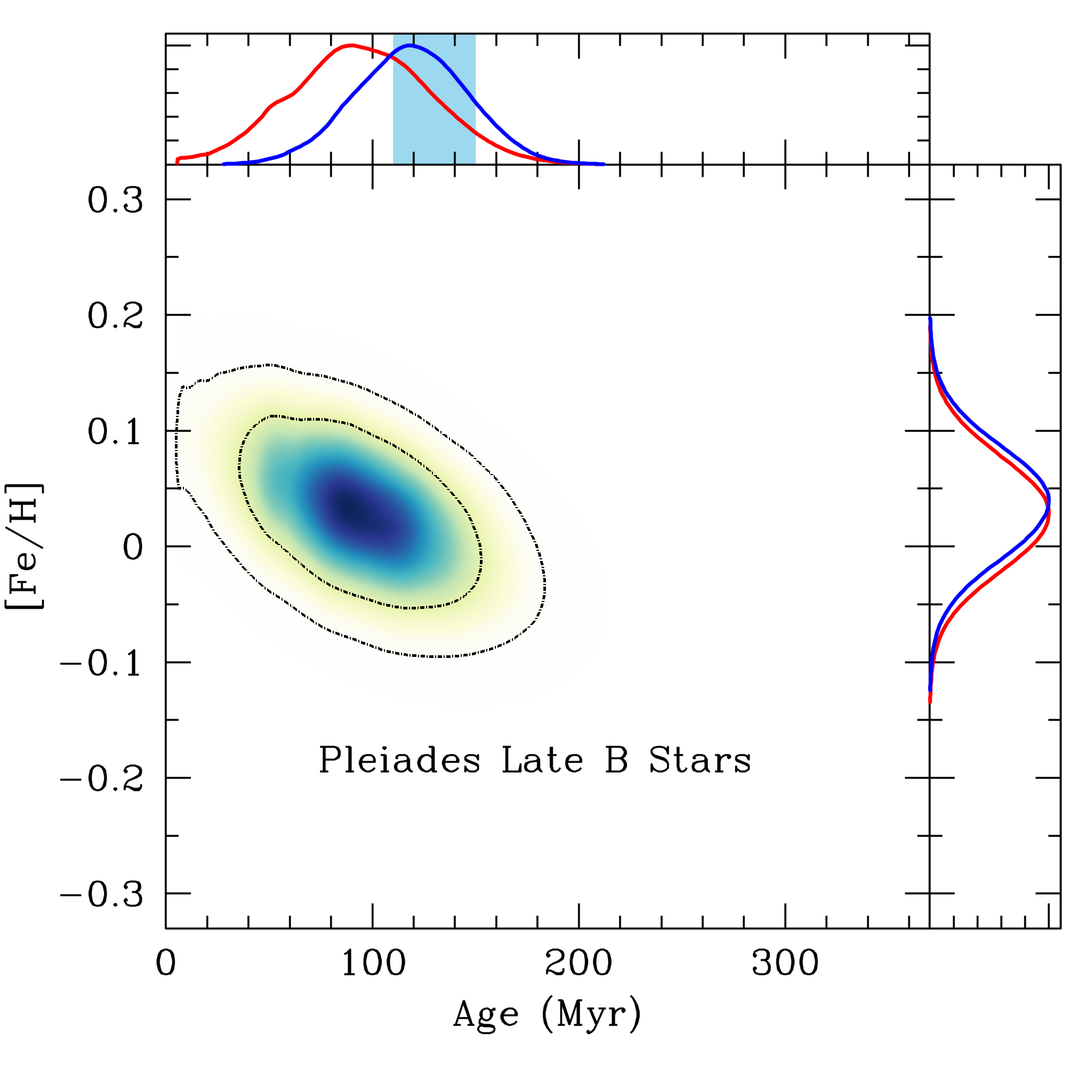}
\caption{Posterior probability distribution in age and metallicity for the Pleiades, measured by fitting six late B stars, HIP 17527, 17588, 17664, 17776, 17862, and 17900, to the stellar models of \cite{Georgy+Ekstrom+Granada+etal_2013} with (red curves) and without (blue curves) allowing for a stellar rotation.  The cluster's age has been measured to be $130 \pm 20$ Myr using the lithium depletion boundary \citep{BarradoyNavascues+Stauffer+Jayawardhana_2004}; this interval is shaded light blue.  The \cite{Georgy+Ekstrom+Granada+etal_2013} give a slightly younger age, albeit still consistent at $1\sigma$.  There is little difference between the age and metallicity derived with and without considering rotation.  As Figure \ref{fig:isochrones} shows, the rotating and nonrotating tracks for $\sim$2--2.5 $M_\odot$ stars only separate appreciably at older ages.}
\label{fig:pleiades}
\end{figure}

Figure \ref{fig:pleiades} shows our results.  With a cluster $[{\rm Fe/H}] = 0.03 \pm 0.05$, the marginalized probability distribution is $\sim 95 \pm 35$ Myr ($1\sigma$), consistent with age determinations using the lithium depletion boundary \citep{BarradoyNavascues+Stauffer+Jayawardhana_2004} or higher-mass turnoff stars \citep{Meynet+Mermilliod+Maeder_1993}.  Fixing the metallicity to $[{\rm Fe/H}] = 0.03$ has little effect on the results, changing the $1\sigma$ uncertainty from $\sim$35 to $\sim$30 Myr.  

The most massive Pleiads, HIP 17499 (Electra) and HIP 17702 (Alcyone), both strongly prefer isochrone-based ages under 100 Myr (assuming the same distance and extinction), with Alcyone fitting $\sim$6.5 $M_\odot$ models at $\sim$70 Myr.  Electra's best-fit age of $\sim$90 Myr is consistent with the other Pleiads.  The \cite{Ekstrom+Georgy+Eggenberger+etal_2012} stellar models at $\Omega_0/\Omega_{\rm crit} = 0.58$ predict no stars $\gtrsim$5 $M_\odot$ near the main sequence: either Alcyone is a blue straggler, or the rotating stellar models fail to match massive B stars.  Despite this caveat, the good agreement of the fit to six late B Pleiads with the cluster's known age lends credibility to our application of the same $\sim$2--2.5 $M_\odot$ models to the Ursa Majoris and Hyades clusters.

\subsection{The Ursa Majoris Moving Group}

The Ursa Majoris moving group (UMa) is a coeval stellar association with an age of $500 \pm 100$ Myr derived from isochrone fits \citep{King+Villarreal+Soderblom+etal_2003}.  As a coeval association, it is also expected to be chemically homogeneous.  We can therefore apply the fitting techniques presented in this paper to early-type UMa stars, jointly constraining the association's age and metallicity.  For this exercise, we choose five early-type stars that are listed as UMa members in \cite{King+Villarreal+Soderblom+etal_2003}: HIP 53910, 59774, 62956, 65477, and 76267.  We exclude HIP 6061, which \cite{King+Villarreal+Soderblom+etal_2003} list as a member but which lies well outside the cluster core and is inconsistent with the other stars' ages.  In the case of HIP 76267, we subtract the flux of its eclipsing binary companion (a G5V star, \citealt{Tomkin+Popper_1986}).  HIP 65477 has an M-type companion \citep{Mamajek+Kenworthy+Hinz+etal_2010} which is too faint to appreciably affect its optical flux.  We exclude the other early-type UMa members from \cite{King+Villarreal+Soderblom+etal_2003}, all of which appear to be spectroscopic binaries with companions of unknown spectral type.  Four of our five stars, HIP 53910, 59774, 62956, and 65477, are members of the UMa nucleus.  

\begin{figure}
\includegraphics[width=\linewidth]{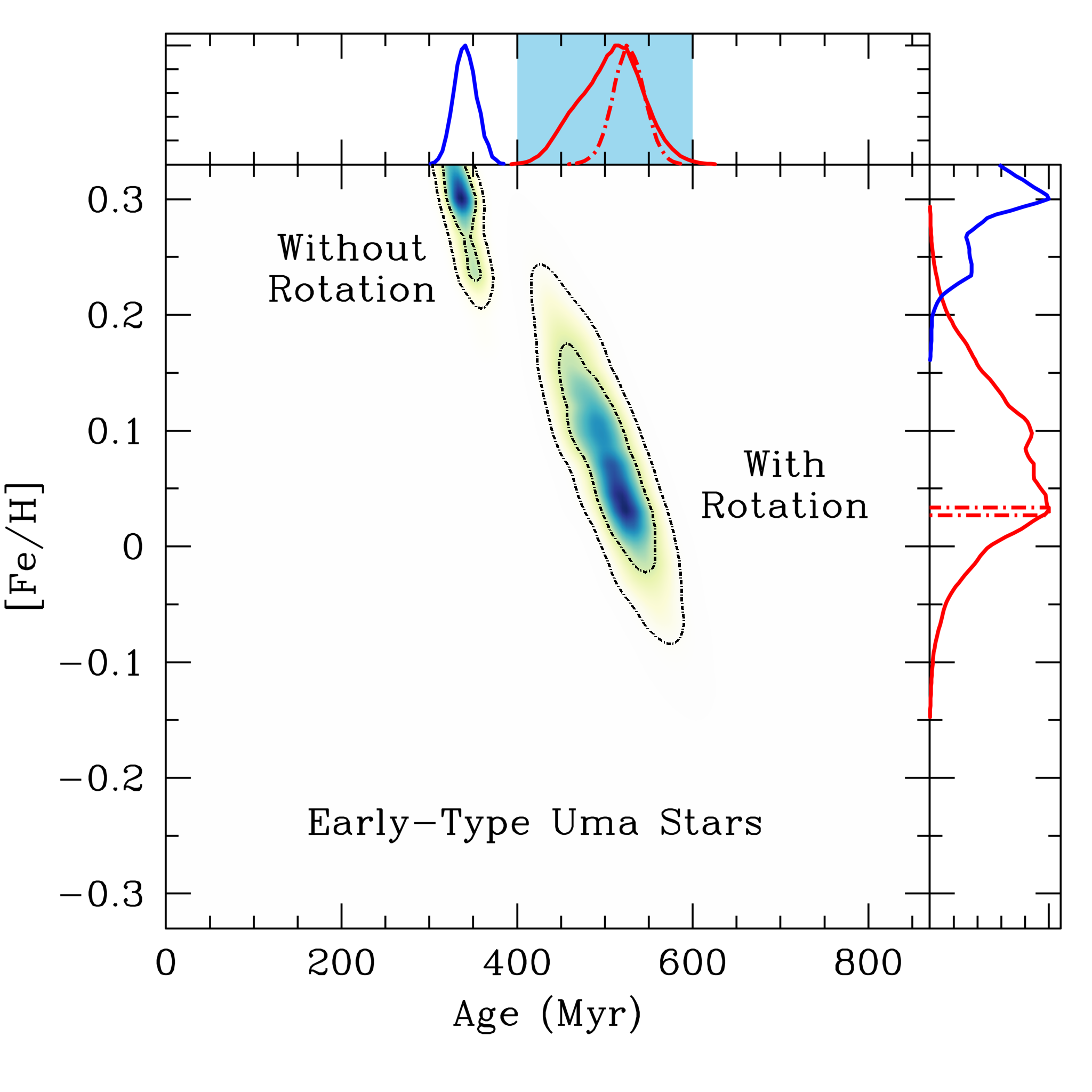}
\caption{Joint posterior probability distribution for the age and metallicity of five early-type Ursa Majoris moving group members, with a metallicity prior of $[{\rm Fe/H}] = 0 \pm 0.1$.  Our derived age with rotation agrees well with $\sim$500 estimated by \cite{King+Villarreal+Soderblom+etal_2003}.  The red dot-dashed curves show the constraints from assuming a delta-function metallicity prior $[{\rm Fe/H}] = 0.03$, the central value measured from FGK candidate UMa members \citep{Tabernero+Montes+Gonzalez-Hernandez+etal_2014}. When neglecting rotation, the age/metallicity constraints for HIP 62956 do not agree with the other stars; multiplying the probability densities yields the implausibly young age and high metallicity shown.  }
\label{fig:uma_tz}
\end{figure}

When fitting the ensemble of stars, we assume a Gaussian metallicity prior for the cluster centered on $[{\rm Fe/H}] = 0$ with a dispersion of 0.1 dex.  We assume the individual stars to have the (unknown) cluster value.  Each individual star may have its own parallax, rotation and orientation; we marginalize over all of these parameters star-by-star.  We then multiply the joint posterior probability distributions in age and metallicity for all stars, obtaining the results shown in Figure \ref{fig:uma_tz}.  The upper-left density plot assume all stars to be nonrotating while the lower-right assumes a Maxwellian prior on $\Omega_0/\Omega_{\rm crit}$.  The inner and outer contours enclose 68\% and 95\% of the probability, respectively.  The dot-dashed curves show the results if the metallicity prior is a delta function at $[{\rm Fe/H}] = 0.03$.  

When using the rotating models, all of the stars are consistent with one another in their age-metallicity constraints; their 1$\sigma$ contours overlap one another.  The derived age is in good agreement with the $500 \pm 100$ Myr reported by \cite{King+Villarreal+Soderblom+etal_2003}.  Our isochrone analysis also favors a very slightly super-Solar metallicity, in agreement with recent spectroscopic measurements of later-type members.  \cite{Tabernero+Montes+Gonzalez-Hernandez+etal_2014}, using a sample of 44 candidate FGK members, find that 29 of the 44 stars share a similar chemical composition, with $[{\rm Fe/H}] = 0.03 \pm 0.07$ dex.  Using a delta function metallicity prior at $[{\rm Fe/H}] = 0.03$ (dot-dashed red lines) gives an age of $530 \pm 40$ Myr ($2\sigma$ formal errors).  

The agreement is much worse when neglecting rotation.  Much of this is due to HIP 62956: its $1\sigma$ contour does not overlap those of the other stars in age-metallicity space.  Excluding this star yields good agreement in the nonrotating case at an age of just over 400 Myr, which is still $\sim$20\% younger than in the rotating case.  Including or excluding HIP 62956 has a negligible effect on the constraints from rotating stellar models.

\subsection{The Hyades Open Cluster}

We now revisit the age of the Hyades open cluster, calculated to be $625 \pm 50$ Myr by \cite{Perryman+Brown+Lebreton+etal_1998} using nonrotating isochrones with convective overshooting.  The \cite{Perryman+Brown+Lebreton+etal_1998} constraint is based mostly on five stars near the main-sequence turnoff without indications of multiplicity: HIP 20542, 20635, 21029, 21683, and 23497.  Reddening between the Sun and the Hyades is negligible \citep{Taylor_2006}.  We use the same five stars in our present framework to place new constraints on the Hyades age, adopting a metallicity of $[{\rm Fe/H}] = 0.10$ \citep{Taylor+Joner_2005} with a conservative Gaussian error of 0.05 dex.  We also add the individual components of the binary HIP 20894 as two additional stars, with a magnitude difference $\Delta V_T = 1.10$ and $\Delta \left(B_T-V_T\right) = -0.006$ \citep{Peterson+Stefanik+Latham_1993}.  As for Ursa Majoris, we assume the stars to share a composition and marginalize over mass, rotation, and orientation separately for each star.

\begin{figure}
\centering
\includegraphics[width=\linewidth]{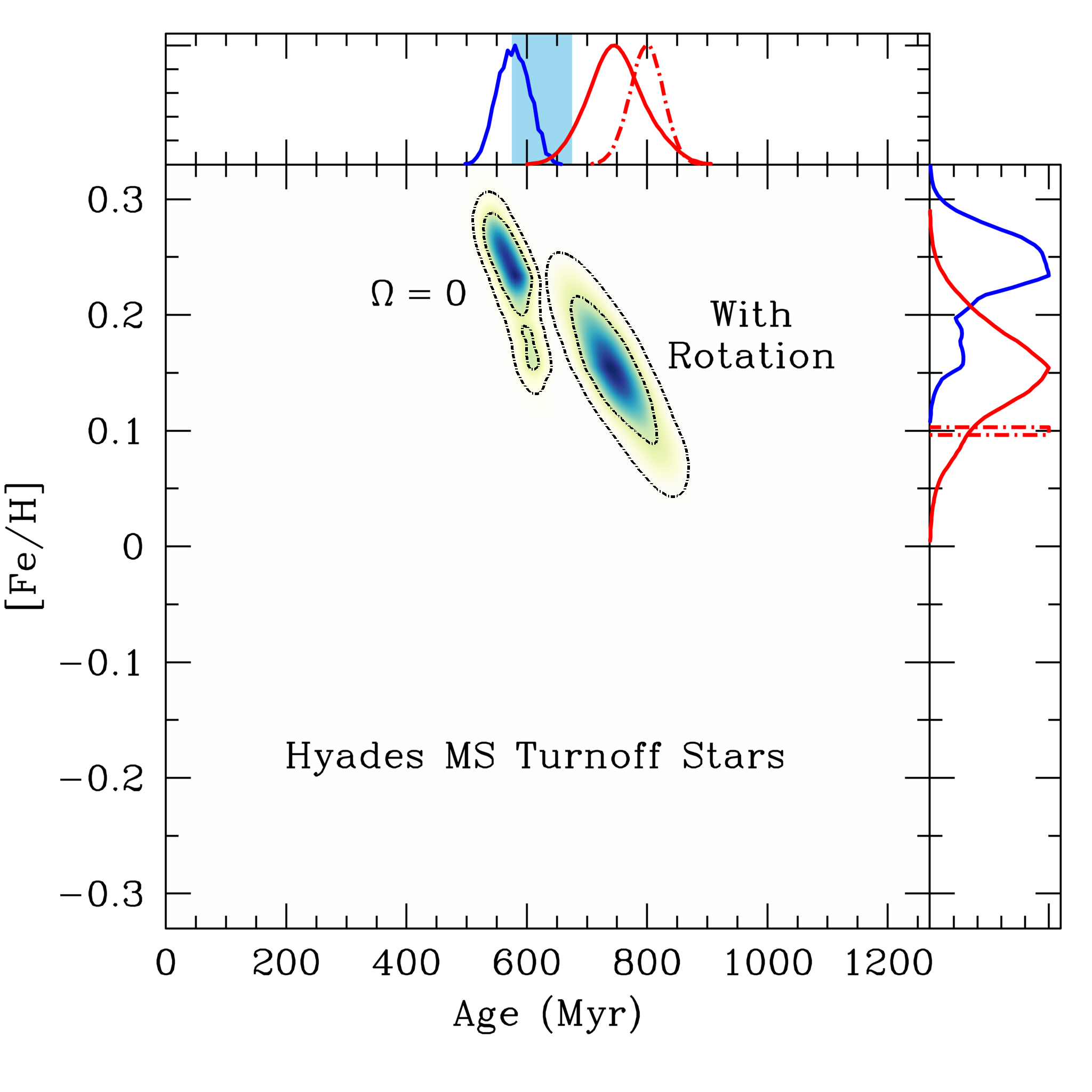}
\caption{Age-metallicity constraints on the Hyades open cluster using seven main-sequence turnoff members (HIP 20542, 20635, 21029, 21683, 23497 and the two components of HIP 20894).  The inner and outer contours enclose 68\% and 95\% of the probability, respectively; the left contours and blue curves neglect rotation.  The best-fit age is $\sim$750 Myr when including an uncertainty of 0.05 in [Fe/H] (solid red line), or $\sim$800 Myr with at $[{\rm Fe/H}] = 0.1$ (dot-dashed red curve).  These ages are significantly older than the consensus age of 625 Myr, which was calculated largely based on five of these seven stars \citep{Perryman+Brown+Lebreton+etal_1998}.  We shade the previous age constraint in light blue for comparison.  }
\label{fig:hyades}
\end{figure}

The lengthening of the main-sequence lifetime with rotation maps onto an older inferred age of $\sim$750 Myr for the Hyades, shown in Figure \ref{fig:hyades}.  As Figure \ref{fig:isochrones} shows, a rotating isochrone can also overlap a slightly more massive nonrotating isochrone near the main sequence turnoff.  Indeed, the difference between the maximum likelihood masses with and without including rotation is typically $\sim$0.05--0.1 $M_\odot$, which corresponds to a $\sim$10\% difference in main-sequence lifespan.  These effects combine to increase the Hyades age from just under 600 Myr to about 750 Myr.  We note that the nonrotating models used by \cite{Perryman+Brown+Lebreton+etal_1998} included slightly stronger core convective overshoot than the \cite{Georgy+Ekstrom+Granada+etal_2013} models ($\alpha = 0.2$ compared to $\alpha = 0.1$).  This mimics some of the effects of rotation, and accounts for the modest difference between the nonrotating models here and in \cite{Perryman+Brown+Lebreton+etal_1998}.

All seven turnoff stars, including the two components of HIP 20894, are consistent with an age of $\sim$750--800 Myr.  The 800 Myr, $\Omega_0/\Omega_{\rm crit} = 0.58$ $Z_\odot$ isochrone in \cite{Ekstrom+Georgy+Eggenberger+etal_2012} has its red clump at $B-V \approx 1$, and $V \approx 0.4$, completely consistent with the location of the four red giants in Figure 21 of \cite{Perryman+Brown+Lebreton+etal_1998}.  The \cite{Georgy+Ekstrom+Granada+etal_2013} nonrotating models strongly prefer a high metallicity for the cluster, significantly higher than the spectroscopic value.  The rotating models are fully consistent with the spectroscopic metallicity; fixing ${\rm Fe/H}] = 0.1$ gives an age of $800 \pm 50$ Myr.

Open clusters like the Hyades are often used as benchmarks to calibrate age dating techniques for lower mass stars.  If the cluster's age really is closer to 800 Myr than to 600 Myr, it may require a re-calibration of secondary age indicators like stellar rotation and activity \citep{Mamajek+Hillenbrand_2008}.  
At the very least, our results suggest that the uncertainty in the Hyades' age is underestimated.  Resolving the discrepancy might require other age indicators for these stars or additional tests of the rotating stellar models.  Asteroseismology of lower-mass stars provides a relatively direct measurement of the central sound speed, and by extension, the mean molecular weight and helium fraction.  The Transiting Exoplanet Survey Satellite \citep[TESS,][]{Ricker+Winn+Vanderspek+etal_2014} could soon provide crucial data for F and G-type Hyades members.  

Two of our seven turnoff stars, HIP 20894 A and B, form a binary with dynamically measured masses.  The most recent and most precise masses are $2.86 \pm 0.06$ $M_\odot$ and $2.16 \pm 0.02$ $M_\odot$ for the primary and secondary, respectively \citep{Torres+Lampens+Fremat+etal_2011}.  If these masses are correct, they raise more questions than they answer:  none of our 2.86 $M_\odot$ stellar models come within 1 mag of the observed luminosity of HIP 20894 A {\it at any age}.  Figure 10 of \cite{Torres+Lampens+Fremat+etal_2011} confirms this, and further shows that 2.86 $M_\odot$ stars with HIP 20894 A's colors are rapidly traversing the subgiant branch.  We note that our analysis, with a $\sim$2.4 or 2.5 $M_\odot$ HIP 20894 A, places the star on the main-sequence turnoff before the subgiant branch, where it is evolving ten times more slowly through color-magnitude space.  Previous orbital solutions of the same system have found very different masses, albeit with much less data \citep{Torres+Stefanik+Latham_1997, Armstrong+Mozurkewich+Hajian+etal_2006}.  A confirmation of HIP 20894 A's high mass would present major challenges to models of stellar evolution.

\section{Conclusions} \label{sec:conclusions}

In this paper, we have used recently computed stellar models with an improved treatment of stellar rotation \citep{Georgy+Ekstrom+Granada+etal_2013} to derive stellar parameters, particularly ages, of nearby early-type stars.  We combine ATLAS9 model atmospheres \citep{Castelli+Kurucz_2004} with the two-dimensional model of $T_{\rm eff}$ and $\log g$ in LR11 to compute synthetic magnitudes as a function of orientation.  We then use fine grids of nonrotating isochrones \citep{Girardi+Bertelli+Bressan+etal_2002} to interpolate between the rotating models in time, mass, and metallicity.  In the case of metallicity, we also use the nonrotating models to {\it extrapolate} the rotating models to super-Solar metallicities.  

We have applied a Bayesian age analysis to obtain posterior probability distributions of the stellar parameters assuming random orientations and a standard initial mass function.  While we have computed synthetic photometry through a wide range of filters, the analysis we present relies only on the {\it Tycho} $B_T V_T$ system and {\it Hipparcos} parallaxes \citep{ESA_1997}.  The photometry and astrometry, which have since been updated \citep{Hog+Fabricius+Makarov+etal_2000, vanLeeuwen_2007}, provide uniform and high-quality data for all bright stars.  

We have applied our analysis first to stars with known ages, including the Pleiades open cluster.  Our use of the new rotating isochrones recovers, within its errors, the known ages of $\beta$ Pictoris and (with caveats) two members of the AB Doradus moving group.  Fitting of the late B Pleiads recovers an age of $\sim$100 Myr, in good agreement with previous results using both isochrone fitting and the lithium depletion boundary \citep{Meynet+Mermilliod+Maeder_1993, BarradoyNavascues+Stauffer+Jayawardhana_2004}.  Fitting the massive Pleiad Alcyone favors a younger age.  This could indicate that Alcyone is a blue straggler, or that the rotating stellar models fail at higher masses.

Fitting $\sim$2--2.5 $M_\odot$ models to the Ursa Majoris moving group agrees well with previously published ages for that cluster\citep{King+Villarreal+Soderblom+etal_2003}.  Fitting the turnoff stars of the Hyades, however, implies a cluster age of $\sim$750--800 Myr, significantly older than the previous estimate of $625 \pm 50$ Myr using the same stars \citep{Perryman+Brown+Lebreton+etal_1998}.  This difference results from an increase in both main-sequence lifetime and luminosity due to rotational mixing.

The older age we derive for the Hyades turnoff stars raises questions about either the cluster's age, stellar modeling, or both.  Including stellar modeling uncertainties could imply a much larger error of at least 100 Myr for the cluster age, which would propagate into secondary age indicators calibrated on clusters like the Hyades.  Resolving these modeling uncertainties will likely require independent age indicators, perhaps asteroseismology of F and G type Hyades stars.

Finally, we provide a web interface at \verb|www.bayesianstellarparameters.info| where a user may obtain the posterior probability distributions of age, mass, and orientation for any bright ($H_P < 9$), blue ($B-V < 0.25$) {\it Hipparcos} star with a good distance ($\varpi/\sigma_\varpi > 15$).  The web server currently assumes negligible extinction and uses only {\it Tycho} $B_T V_T$ photometry and {\it Hipparcos} parallaxes.  A careful analysis would be needed to place other photometry, like 2MASS, on the same absolute scale, and ensure that the resulting ages remain consistent with one another and with independently determined ages.

\acknowledgments{The authors would like to thank the computing staff at the Institute for Advanced Study for help in setting up a web interface, Waqas Bhatti for the server script
template, and Josh Schlieder, Lynne Hillenbrand, and Geza Kovacs for helpful comments on the manuscript.  The authors thank the referee, Juan Zorec, for helpful comments and for pointing out an error in the age scale.  This work was performed in part under contract with the Jet Propulsion Laboratory (JPL) funded by NASA through the Sagan Fellowship Program executed by the NASA Exoplanet Science Institute.  This research has made use of the SIMBAD database and the VizieR catalogue access tool, operated at CDS, Strasbourg, France.}

\bibliographystyle{apj_eprint}
\bibliography{refs}

\end{document}